\documentclass[sigplan,nonacm,citsort,urlbreakonhyphens=true]{acmart}
\usepackage{graphicx} 

\usepackage{xurl}
\hypersetup{breaklinks=true}
\PassOptionsToPackage{hyphens}{url}

\usepackage{ifthen}
\def\oldversion{oldversion}

\title{How to Secure Existing C and C++ Software \\ without Memory Safety}
\author{ {{\Large \'Ulfar Erlingsson}}%
\raisebox{-0.5ex}{\textsuperscript{*}} 
}
\affiliation{ 
              \city{} \country{} }

\date{March 2025}

\begin{document}

\begin{abstract}
\ifthenelse{\equal{\oldversion}{newversion}}{%
} 
{ 
The most important security benefit of software memory safety is easy to state: 
for C and C++ software, attackers can exploit most bugs and vulnerabilities 
to gain full, unfettered control of software behavior, 
whereas this is not true 
for most bugs in memory-safe software---just a few~\cite{Log4J}.

Fortunately, this security benefit---most bugs don’t give attackers full control---can 
be had for unmodified C and C++ software, 
without per-application effort. 

This doesn’t require trying to establish memory safety; 
instead, it is sufficient to eliminate most of the combinatorial ways 
in which software with corrupted memory can execute. 
There already exist practical compiler and runtime mechanisms to
eliminate these avenues for attack; these mechanisms
incur little overhead 
and need no special hardware or platform support.

Each of the mechanisms described here is already in
production use, at scale, on one or more platforms.
By supporting their combined use in development toolchains, 
the security of all C and C++ software 
against remote code execution attacks
can be rapidly, and dramatically, improved.
}
\end{abstract}

\maketitle

\renewcommand{\thefootnote}{} 
\footnotetext{
\raisebox{-0.5ex}{\large *}~The opinions expressed here 
are those of the author, not their employer.
}

\section{Motivation}
In modern computing, 
attackers continue to exploit memory-corruption vulnerabilities 
to devastating effect.

The risk is especially high for server-side software, as software on client devices, such as smartphones, has been significantly hardened against attacks~\cite{zeroday}.  
On servers,
most of the foundational platforms and libraries 
are written 
in C and C++, where memory-corruption is possible.
For high assurance,
these foundations must be rewritten 
in memory-safe languages like Go and Rust~\cite{CISA};
however, history and estimates
suggest this will take a decade or more~\cite{Rebert}.

There is a current, urgent 
need to make rapid progress on reducing 
the risk of attacks in 
existing C and C++ software.
In this, we must temper our expectations: 
without memory safety,
attackers will still be able to change software behavior, 
at least somehow.
However, it should be possible to prevent attackers 
from completely controlling software behavior---at
least in most cases, for the most serious types of attacks---even
when attackers can exploit memory-corruption bugs.
For this goal,
the recent improved security outcomes
for client software
can give us hope~\cite{apple,zeroday},

\section{A Pragmatic Security Goal}
Remote Code Execution (RCE) attacks---where
attackers exploit memory-corruption bugs
to achieve complete control---are
a very important class of potentially-devastating attacks.
Such attacks can be hugely disruptive, 
even simply in the effects and economic cost of
their remediation~\cite{Log4J}.
Furthermore, the risk of such attacks
is of special, critical concern
for server-side platform foundations~\cite{CISA}.

Greatly reducing the risk of RCE attacks in C and C++ software,
despite the presence of memory-corruption bugs,
would be a valuable milestone in software security---especially
if such attacks could be almost completely prevented.

We can, therefore, aim for the ambitious, pragmatic goal of
preventing most, or nearly all, 
possibilities of RCE attacks 
in existing C and C++ software without memory safety.
Given the urgency of the situation,
we should only consider 
existing, practical security mechanisms
that can be rapidly deployed at scale.

\section{Software Security without Memory Safety}
There are a number of
software security mechanisms
that are already implemented on major platforms
and have seen widespread deployment and use.

From these, there are several
practical, low-overhead techniques 
known for
preventing attacker exploits 
of memory-corruption bugs in existing C and C++ software.

Finally, from those, 
there are a few
that require neither 
special operating-system or hardware support
nor any significant per-application effort.

Below are four such exploit-mitigation technologies.

\paragraph{Stack Integrity}
Attackers delight in memory-corruption bugs
that allow them to modify or control 
one of the function-call stacks of executing C and C++ software,
since this can allow them to completely dictate behavior~\cite{Hovav}.


Two decades ago, 
CCured and XFI showed 
how each C and C++ execution stack 
can be isolated from the effects of memory-corruption bugs,
effectively or absolutely,
in an efficient manner~\cite{CCured,XFI}.
For this,
compilers need only generate code
as if for a segmented-memory system~\cite{SSeg},
permitting only constant updates to the stack pointer
and moving all pointer-accessed variables off the stack---this
can be done 
efficiently using a dedicated, thread-local heap region~\cite{XFI}.

Such \textit{stack integrity} 
is already implemented by the LLVM compiler, 
in a practical manner~\cite{SStack}.
Its low overhead can be made negligible 
(by static analysis of C++ references),
and its guarantees can be made absolute
by preventing writes through pointers 
from modifying any stack memory.

With stack integrity,
attackers cannot change 
the arguments, local variables, or return sites of functions---as
long as function calls are also restricted,
e.g., as described next.

\paragraph{Control-Flow Integrity}
To gain complete RCE control,
attackers must direct execution to code of their choosing.

It is simple to block attackers from executing their own code,
e.g., by preventing the addition of any new code.

However,
attackers may still
exploit memory corruption to
redirect execution within the existing software code~\cite{Hovav}.
Fortunately, 
in C and C++ software---once stack integrity is enforced---any
such redirection can be tamed by
preventing unintended use of
function pointers and C++ vtables.

\textit{Control-flow integrity} (CFI)
ensures that each
function-call site will---despite any memory corruption---only 
ever direct execution to
the start of a compatible function.
Specifically,
control-flow integrity can ensure that any function called via a pointer
has a type signature,
or is a (virtual) member function of a class,
that matches the source code for the call site~\cite{CFI,CompCFI}.

Control-flow integrity is widely enforced by low-overhead,
compiler-added checks in both LLVM and gcc~\cite{llvmCFI,CompCFI},
as well as in Windows compilers~\cite{WinCFG}, 
sometimes with hardware support~\cite{x86CFI,armCFI}.
Windows~10~\cite{winCFI}, Android~\cite{droidCFI},  
and the Chrome Web browser~\cite{chromeCFI} 
all enforce 
a form of control-flow integrity in their production software.

Combined with stack integrity,
control-flow integrity can inductively guarantee
that software always executes, as intended,
as a well-nested sequence of functions
with uncorrupted arguments and variables~\cite{XFI}.


%
%
\paragraph{Heap Data Integrity}
Attackers
usually must perform ``\textit{heap feng shui}''
(i.e., manipulate the precise heap layout)
to be able to corrupt heap memory
in a targeted way~\cite{heapFS}. This corrupted
heap memory can then be used to redirect
software control flow via function pointers
and C++ vtables.

There are many efficient ways to 
improve the integrity of the C and C++ heap,
most using hardware support~\cite{MTE,apple}.

One simple software means of protecting the heap
is to partition memory into many disjoint regions,
such that each static heap-object allocation site 
uses only one region~\cite{SafeCode}.

With enough partitions,
attackers lose their
 ``feng shui'' control over heap layout,
and thereby nearly all means
for targeted heap corruption---whether 
by temporal bugs, like use-after-free, 
or spatial bugs, like overflows.

These benefits are the strongest, and most clearly seen,
when a dedicated region is assigned to each heap allocation site,
and each partition 
contains only heap objects of the same type and origin.
Furthermore, 
with careful assignment to regions,
the attackers' ability to exploit memory corruption
can be reduced even further
by confining each use of a heap pointer 
to the range of heap partitions that are
statically accessible 
at the usage site in the source code~\cite{SafeCode,DFI}.

Attackers are impeded even by coarse
heap partitioning, such as by size~\cite{tcmalloc}.
However,
stronger \textit{heap integrity}
can be had by partitioning more finely,
using static 
namespaces, types, etc.
This is increasingly done,
as it allows trading 
higher resource overheads for improved security, for
instance in the Chrome Web browser~\cite{PAlloc} 
and Apple's iOS/MacOS~\cite{kallocType},
both of which ship with partitioned allocators.


\enlargethispage{.5\baselineskip}

\paragraph{Pointer Integrity and Unforgeability}
To be successful,
RCE attacks using memory corruption
must, near always, reliably retarget some pointer.
Such retargeting can be made next to impossible,
by making C and C++ pointers 
more similar to true capabilities---that is
\textit{unforgeable} and \textit{unguessable}.

\enlargethispage{.5\baselineskip}
Randomly-generated secrets can be used to 
make pointers emulate true capabilities,
especially on 64-bit hardware.

Randomizing the layout of memory with a secret
is a common RCE defense,
especially useful on servers,
as it forces attacks to take an often difficult derandomization step~\cite{ASLR}.

Each pointer value can also be randomized,
using a different secret for each type of pointer,
as long as pointers are derandomized in the software code before each use.

This is effectively what is done in Apple software, 
which
uses special ARM hardware support
to also check \textit{pointer integrity} at runtime---i.e., ensure
each pointer access
uses pointers of the right type~\cite{apple,PtrAuth}.
Apple uses this further
to enforce a form of stack integrity, 
control-flow integrity, and heap integrity~\cite{apple},
and this is supported in the open-source LLVM compiler~\cite{llvmPAuth}.

When used  
as an orthogonal layer of defense, in addition to the 
above three types of integrity, 
pointer randomization---even without runtime checks---is
a formidable barrier to RCE attacks
based on C and C++ memory corruption.

\vspace*{.5ex}
\noindent
\textbf{A Note on Confinement.~}
The above protections, in combination, can prevent 
complete software control by attackers able to corrupt memory.
This is different
and complementary 
to the use of sandboxing and virtualization
to confine software that can be expected to 
exhibit bad behavior~\cite{confine}.
%

%
%
%
\section{Expected Security Benefits}
These four types of integrity,
do not establish memory safety, but merely
attempt to contain the effects of its absence;
therefore, 
attackers will still 
be able to change software behavior
by corrupting memory.
Given this,
will the combination of the above four protections 
be enough to 
significantly reduce the risk of RCE attacks 
in C and C++ software?

Intuitively, we should expect it to do so.
From the attacker's viewpoint,
RCE attacks 
are enabled by the vast number of otherwise invalid code paths
that become possible in the
``weird machine'' of corrupted software execution~\cite{Dullien}.
These protections block all the major mechanisms that attackers use
for hijacking software control flow, including attacking
the return pointer, overwriting function pointers, and
attacks based on virtual function calls.

With these protections,
attackers can choose only new behaviors
that conform to valid, well-nested sequences 
of calls to compatible functions in the software---with
uncorrupted arguments and local variables---that operate 
on partitioned data objects
accessed via almost-unforgeable pointers.

This intuition is borne out by experience: in part
as a result of Apple's deployment of these defenses since 2019,
the incidence of RCE attacks on Apple client software
has decreased significantly---despite strong attack pressure---and 
the market value of
such attacks risen sharply~\cite{zeroday}.

Of course, 
attackers may still be able to achieve RCE
due to factors
such as misconfiguration or bugs in the protections, platform, or hardware.
However, 
such factors  
are always a risk with any protection mechanism, 
even software memory safety.
As long as there are means to address such issues, if they occur,
security can still be maintained---e.g.,
as demonstrated by hypervisor bugs~\cite{xbox},

More disconcertingly, in some cases,
attacks may be still be possible despite
the above protections.
For example, 
in software that contains a general execution engine, such as a
JavaScript interpreter, 
an RCE attack may be possible by corrupting the interpreter inputs
or by
exploiting logic defects in the engine itself.
Often, as in this example,
similar attacks might also be possible in software
(e.g., a JavaScript engine) written in a memory safe language---however,
we should not assume that this will always be the case.

We should prepare for a case 
where these defenses will fail to 
protect against a specific vulnerability 
in some specific software:
attackers are highly skilled and innovative~\cite{JBIG},
and have full access to modern tools,
like SMT constraint solvers and AI models~\cite{smt}.
This said, we can also expect any systematic weaknesses
to be quickly identified:
the worldwide community of low-level offensive
security researchers has strong motivation 
to uncover flaws in widely-used protections,
and thereby contribute to their improvement.
Any such cases would not refute 
the benefits of the above protections---at least not
if we are prepared
and can address whatever gap was revealed in the defenses.

In all cases, the above protections 
form a sound basis for whatever additional 
defenses are needed,
whether it be the isolation of JavaScript interpreter inputs
in dedicated heap partitions
or the addition of further runtime checks.
For example, together,
stack and control-flow integrity
are sufficient to guarantee
that the code generated for any checks 
immediately before a system-call instruction
will execute---with uncorrupted local variables---whenever,
if-and-only-if
the system-call instruction executes~\cite{CFI}.
These guarantees are even strong enough
to securely corral the execution of 
just-in-time-compiled and self-modifying code~\cite{SelfMod}.

Most importantly,
even if a few gaps appear that need to be shored up,
we can expect
that the above four types of protection---when 
properly implemented and deployed---will
fulfill our pragmatic ambitions,
and prevent nearly all, 
possibilities of RCE attacks 
in existing C and C++ software without memory safety.

\section{Improving C and C++ Security at Scale}
The four protection mechanisms discussed above
are all in production use, in one way or another,
at scale, in widely-used software.
However, their use is the exception, not the rule,
and their use---in particular in combination---requires
security expertise and investment 
that is not common.
%
%
For them
to provide real-world, large-scale improvements 
in the security outcomes
of using C and C++ software,
there remains significant work to be done.
In particular, 
to provide security benefits at scale, for most software,
these protections must be made an integral, easy-to-use
part of the worldwide software development lifecycle.
\vspace*{.75ex}

\noindent
This is a big change and will require a team effort:

\textbf{Researchers and standards bodies} need to work
together to define a set of protection profiles that can be
applied to secure existing software---without new risks or difficulties---easily, at the flip of a flag.

\textbf{Software toolchain developers} need to update their
tools to fully support these protections and---to facilitate adoption---add optimizations that ensure their security benefits
can be had at the smallest possible cost,
in terms of performance and resource overhead.

This work must consider
aspects such as
support for dynamic libraries, pre-compiled object modules, 
and debugging.
Compilers like LLVM
have shown how these aspects can be addressed~\cite{llvmCFI},
but there will still be difficulties to overcome---e.g., 
in compilation for legacy embedded systems---although
the security benefits
may be especially high
in those cases.

\textbf{Application developers} need to configure their 
toolchains to use the new protections (if not enabled by
default), and deploy the more secure software code that results.
\vspace*{.75ex}

\noindent
The current, high-risk state of C and C++ software security
developed over many decades, and will not improve overnight.
The good news is that---as described above---we
can greatly improve the security outcomes for such software,
without this taking further decades.

By changing a relatively small number 
of software development toolchains,
and encouraging their use,
the security of most unmodified, existing C and C++ software 
can be greatly, fundamentally improved, in a practical manner,
at a very rapid rate.

\bibliographystyle{ACM-Reference-Format}
\bibliography{main}

\end{document}